# Reference Ultra Low DC Current Source (ULCS) Between 1 fA and 100 pA at TÜBİTAK UME

Ömer ERKAN, Yakup GÜLMEZ, Cem HAYIRLI, Gülay GÜLMEZ, Enis TURHAN

*Abstract*— **In this paper, we present a programmable Ultra Low DC Current Source (ULCS) developed at TÜBİTAK UME. The output current range is from 1 fA up to 100 pA with 100 aA resolution and is directly traceable to DC voltage, capacitance and time units. The principle of the device is based on applying a linear ramp voltage on standard capacitors with values between 1 pF and 1000 pF. These standard capacitors are commercial standard capacitors which are kept in a temperature controlled insulated box. Linear ramp voltage is generated using a commercial DAQ card. Current source device is fully automated with computer control and can be used to generate any currents with 100 aA resolution in its full range. The uncertainty is 2.5 mA/A at 1 fA current.**

## I. Introduction

Generation of low direct currents is of great interest for the calibration of DC low current meters and low current detectors. One possible way of generating current is based on applying voltage on a standard resistor. Current sources based on high value resistors are not suitable for generating currents lower than pA level because of resistor instability and high temperature dependence of resistors. To accomplish current sources in the sub-picoampere levels, capacitor charging method is more suitable than applying voltage on a standard resistor [1]. In the capacitor charging method, a linear voltage ramp is applied on a capacitor to obtain stable DC low currents. It is possible to obtain ramp voltage with different techniques. One way of obtaining voltage ramp is based on analog integrators which needs for compensation of non-linearities of the integrator circuit [2]. The other way is to use digital/analog (D/A) converters with the compensation of the D/A converter errors [3], [4], [5].

We preferred to use a commercial NI (National Instruments) D/A card with 24 bit resolution to generate voltage ramp. In our method we do not make any corrections to the output of the D/A converter. The non-linearity of the D/A converter is included in the uncertainty budget in section IV. We used commercial HP 16380A air type standard capacitors in the system.

This paper describes the characteristics of the developed ULCS in details [6].

## II. ULCS Device Parts

ULCS device consists of a ramp voltage generator, temperature controlled capacitor unit, microprocessor unit and electronic temperature control units. These units are housed together into a 19” rack cabinet system.

### A. Standard Capacitors

HP 16380A type standard air capacitors were used as the capacitance standards. These standard capacitors have temperature coefficients about 30-40 (µF/F)/°C according to the manufacturer specifications. This value is an important parameter in the current uncertainty. In order to eliminate the temperature effect, we decided to place the capacitors into a temperature controlled environment. We designed a two layer construction consisting of two aluminum boxes. Flexible heaters were mounted to the inner walls of the inner aluminum box shown in Fig. 1 (a). Four standard capacitors with values of 1 pF, 10 pF, 100 pF and 1000 pF were fitted to the inner aluminum box as shown in Fig. 1 (b). They were fitted very tightly so that the capacitors do not move or fall when the box is turned upside down. The inner box with heaters and capacitors were placed into a bigger size aluminum box which is called outer box as shown in Fig. 1 (c). Between the inner and the outer box there is a space that was filled with Styrofoam for thermal isolation. By this way, fluctuations in ambient temperature did not affect the temperature in the box and the temperature of capacitors became more stable with th. There is a small gap between the connection terminals of standard capacitors and the cover of the inner box. Two NTC sensors were used for measuring temperature: one for electronic circuit of temperature control unit as feedback sensor, one for monitoring the internal capacitor temperature during the measurements. These sensors were mounted underneath the top cover of the inner box.

### B. Temperature Control Unit

We used an analog proportional-integral (PI) electronic control circuit to keep the temperature of the standard capacitors around 36 °C. We chose the control temperature above the room temperature to decrease the effect of environment temperature changes. In the temperature control electronic circuit, we preferred to use analog feedback which takes samples from the heater power instead of using Pulse Width Modulation (PWM) technique which may introduce some extra noise into the current measurements.

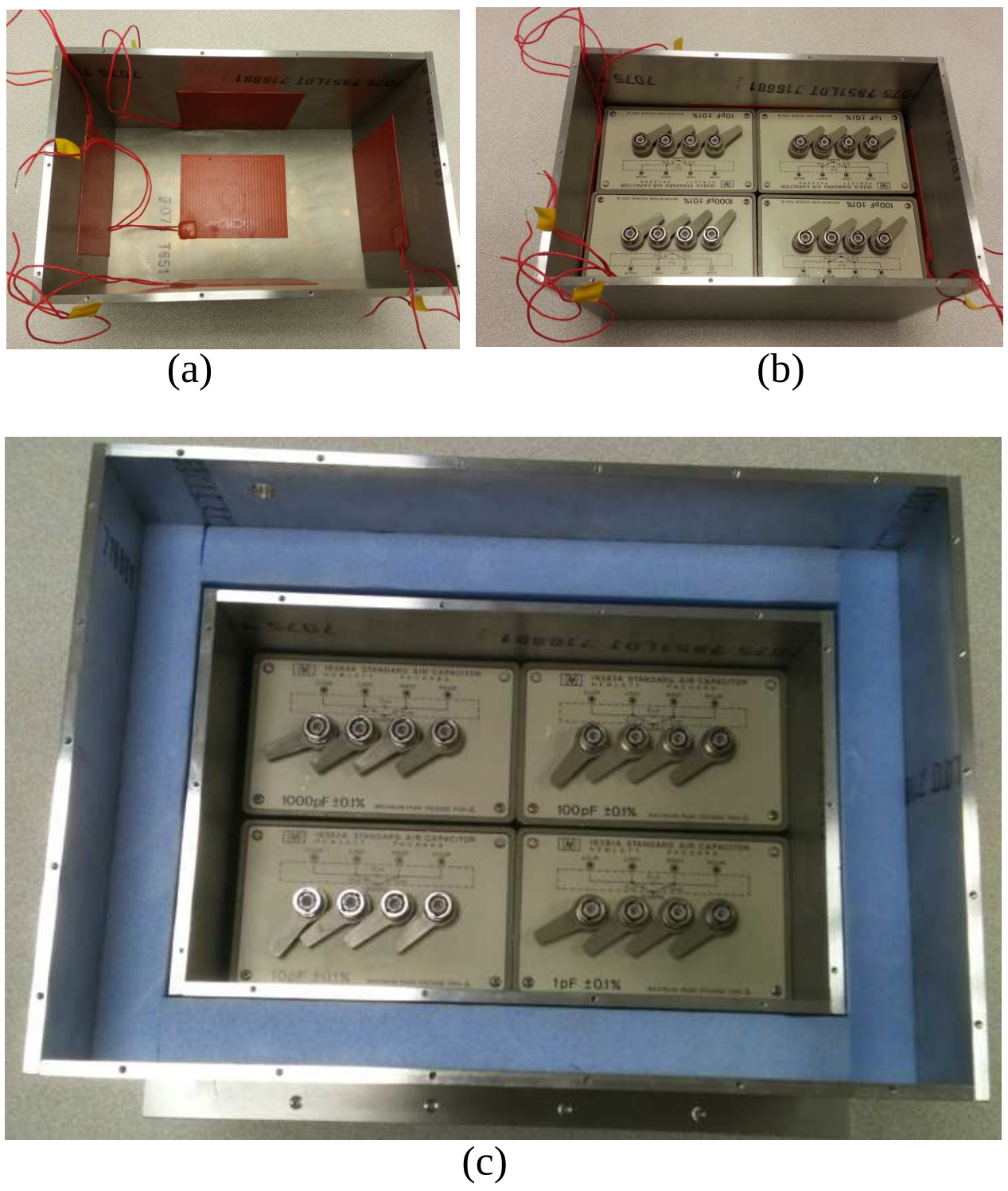

Fig. 1. (a) The flexible heaters were mounted on the inner walls of the inner aluminum box. (b) The standard capacitors were fitted into the inner aluminum box. (c) The inner and the outer boxes were separated by Styrofoam.

We used glass sealed NTC sensors which were previously characterized. Characterization of the NTC sensors was performed in an oven using a Fluke 1594A Super-Thermometer. Resistance changes against temperature were defined and a logarithmic fit were applied to find the temperature values from resistance readings. It was found that the 0.4 Ω change in the resistance value of the NTC sensor corresponds to 1 mK temperature change.

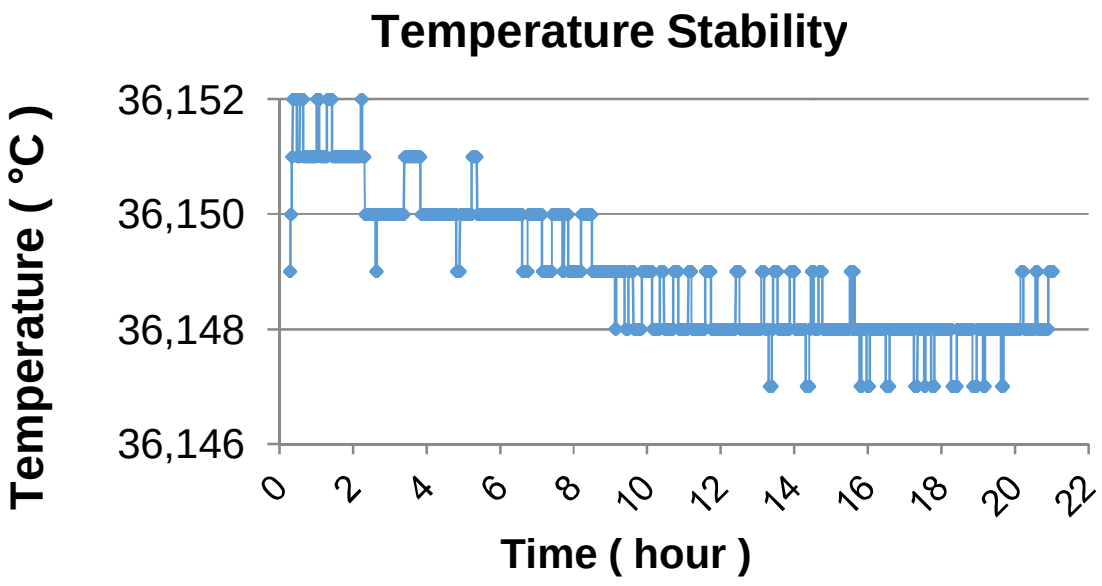

Fig. 2. Temperature stability of standard capacitors.

Measurements were performed with HP 3458A in 4-terminal resistance mode and converted to temperature according to the characterization fit equation of the used NTC sensor.

Temperature stability of the standard capacitors is better than 5 mK/day as can be seen from Fig. 2.

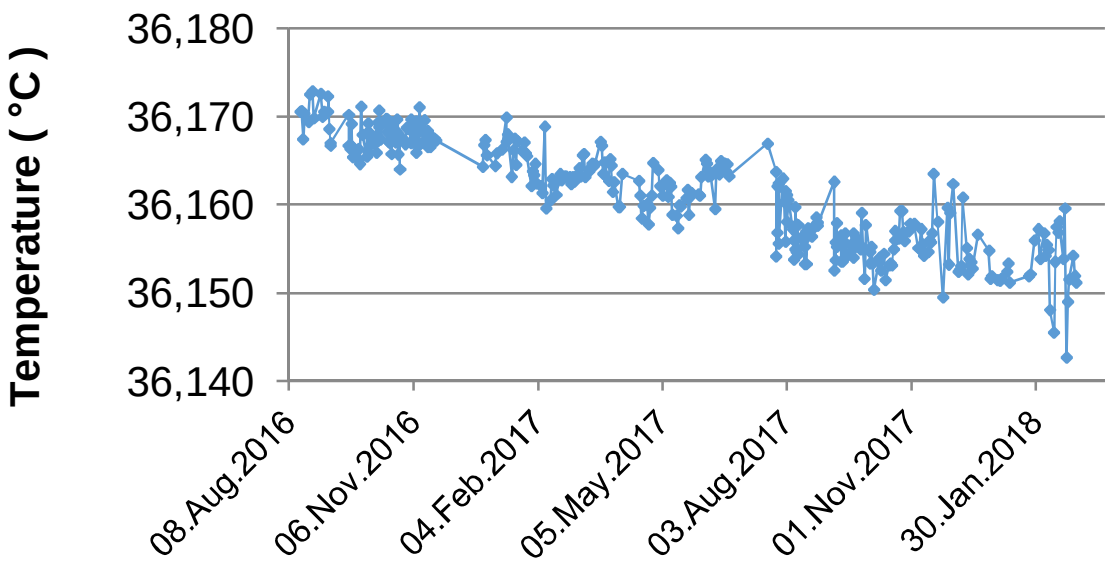

Fig. 3. Monitored temperature of the standard capacitors versus time.

Reproducibility of the temperature of the temperature controlled box is better than 20 mK even though the device stays powered off for long time periods. Fig. 3 shows the long term temperature fluctuations of the standard capacitors over a period of about 1.5 years. As can be seen in Fig. 3, the temperature drift is about 20 mK for about 1.5 years. The stabilization time of the temperature controller is about 4-5 hours after the device is switched on.

### *C. Relay Card*

We designed a relay card to select capacitors automatically for full automatic measurements. Our design criterion for the relay card was not to affect the loss factor of the standard capacitors. The design has a negligible effect on the loss factors of the standard capacitors except 1 pF capacitor. The contact resistances of the relays in the open state are not ideal and become effective in 1 pF range. Dissipation factor of 1 pF capacitor with relay connection is a few parts in $10^5$ while that of 1 pF capacitor without relay connection is a few parts in $10^6$. Since 1 pF is used only for currents lower than 10 fA, this change has a small effect on the produced current value in comparison with the other uncertainty parameters given in section IV.

### *D. Capacitance Measurements*

Capacitance measurements at 1 kHz were performed with an AH2500A capacitance bridge. In addition to the capacitance value, temperature of the capacitors were also measured and recorded regularly.

Measurement results of the capacitors are given in Fig. 4 to Fig. 7. Standard deviations of the fluctuations for all the standard capacitors are around 10 µF/F for more than one year. The large fluctuations seen in the capacitor measurements (for example May. 17, Sep. 17) usually occur when the device switched on after long holiday periods. Each measurement points are average of 10 measurements with standard deviations about 0.5 - 1 µF/F except for 1 pF measurements whose standard deviation is about 5-8 µF/F.

Long term behavior of the standard capacitors seems to be nearly the same. We have investigated the reasons of this correlation. According to the performance checks performed by using Fused-Silica reference capacitors, the fluctuations are too big to be caused by the capacitance bridge. Possible reasons of this behavior may be:

- Hysteresis effect of the capacitors due to large temperature changes when the ULCS switched off
- Humidity changes in the capacitors when the ULCS switched off

There may also be other reasons for this behavior; however, we could not exactly determine the reason.

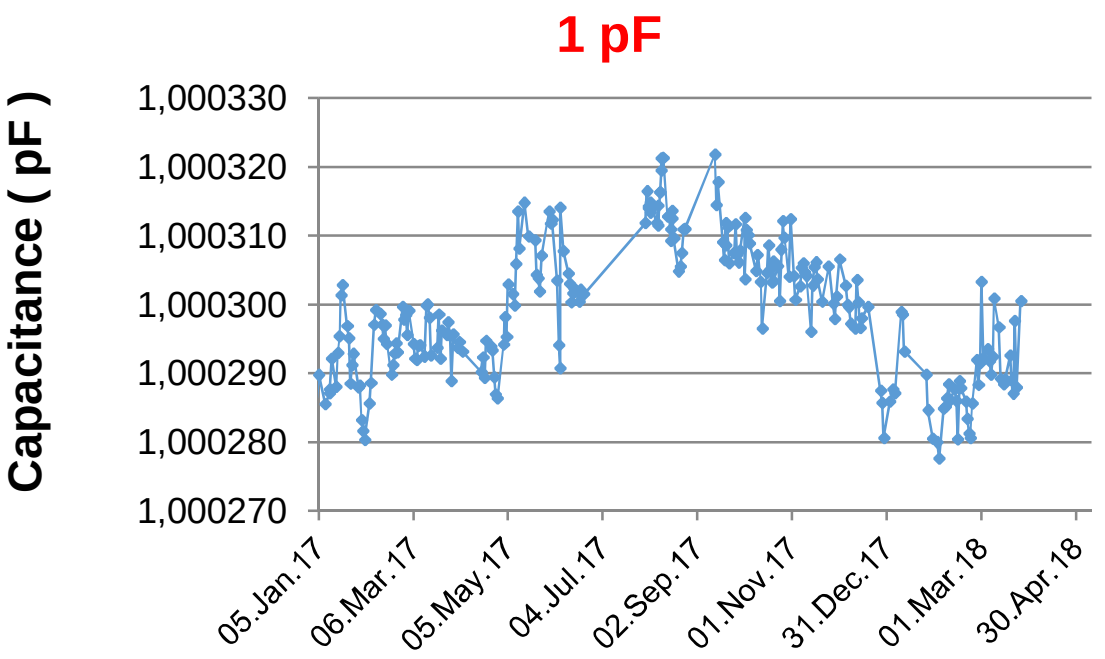

Fig. 4. 1 pF capacitor measurements at 1 kHz.

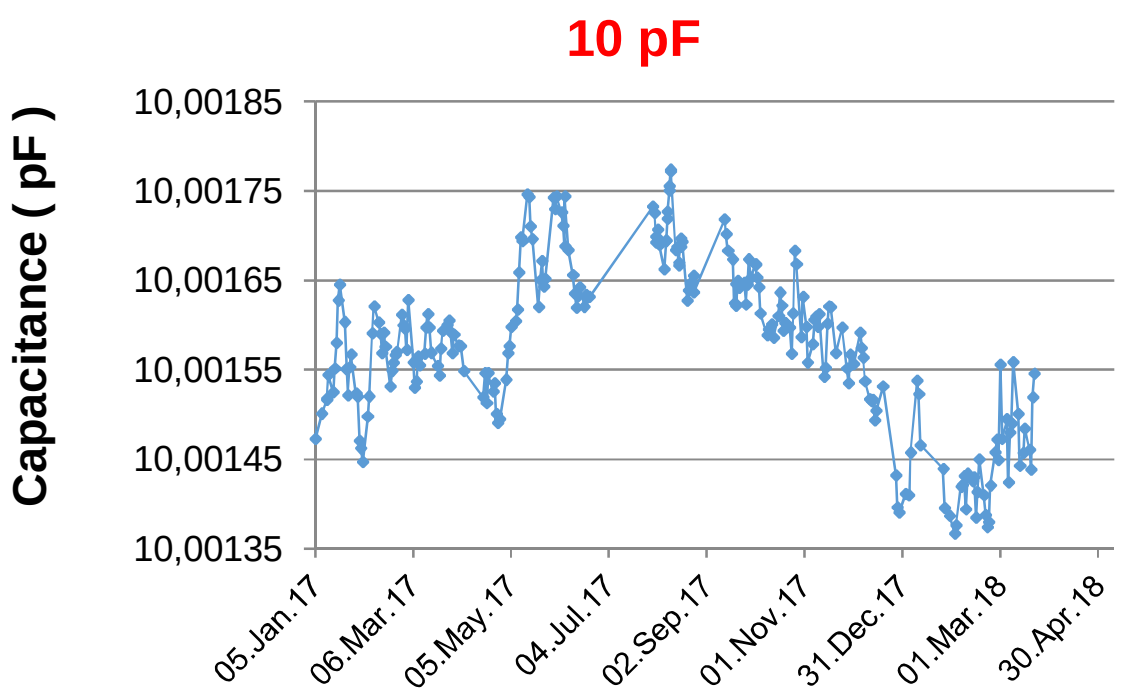

Fig. 5. 10 pF capacitor measurements at 1 kHz.

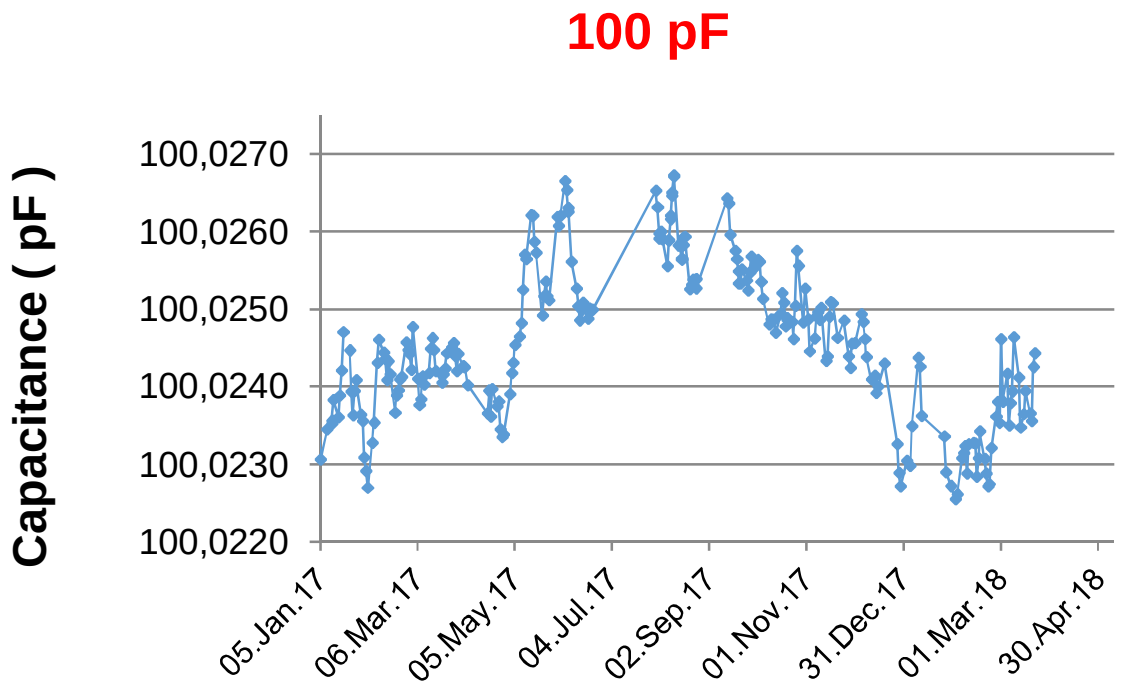

Fig. 6. 100 pF capacitor measurements at 1 kHz.

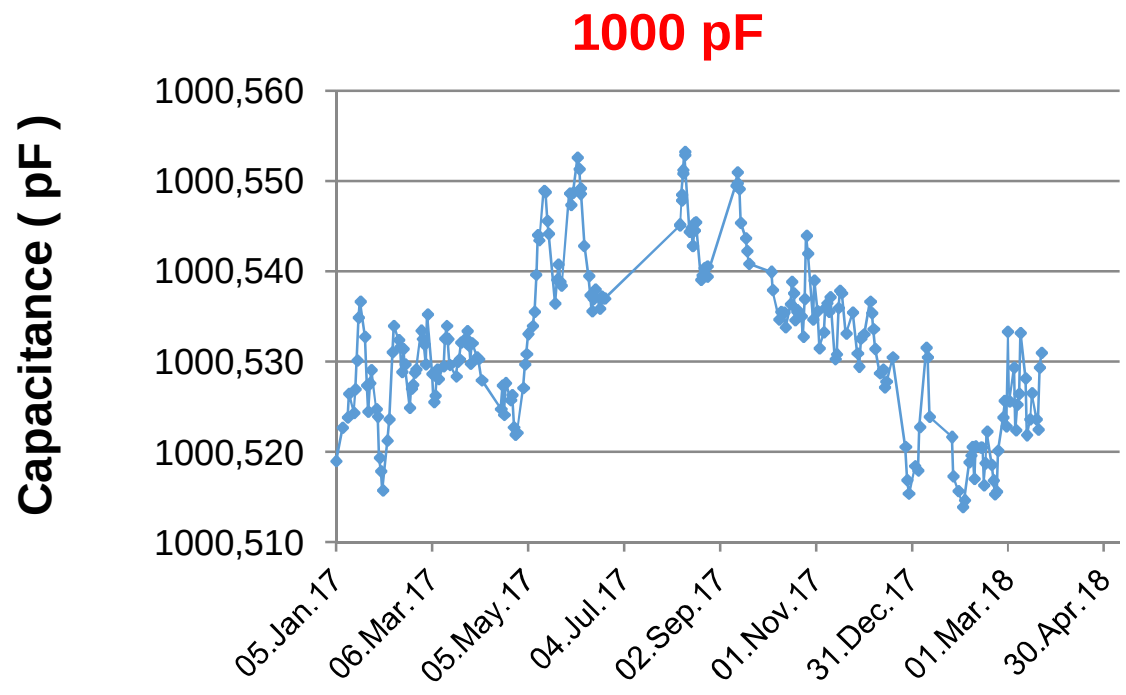

Fig. 7. 1000 pF capacitor measurements at 1 kHz.

### *E. DC Capacitance Measurements*

HP 16380A type standard capacitors used in the ULCS may have large frequency dependence which should be taken into account [7], [8].

We measured the DC capacitance values of 1000 pF and 100 pF. DC capacitance values of 1 pF and 10 pF values were not measured because these capacitors were used for the currents lower than 100 fA. In this current region the AC-DC difference of the capacitors is a negligible parameter in comparison with the other uncertainty parameters. We used a Keithley 6430 ammeter as a transfer standard to compare the current from the ULCS with the current from a resistor-based current source.

In order to obtain DC capacitance value of 1000 pF capacitor, we used 100 pA range of Keithley 6430. The measurement error of the Keithley 6430 electrometer at 100 pA range was determined with an independently generated dc current, using a voltage source of 0.1 V and a calibrated 1 GΩ standard resistor with 25 µA/A uncertainty ($k$=2). Immediately after these measurements, we applied 100 pA current from ULCS to the electrometer and measured the current with the electrometer in the same range. We used 1000 pF capacitor and 100 mV/s ramp voltage in the ULCS in order to obtain 100 pA current value. Using the previously determined measurement error of Keithley 6430 at 100 pA range, we corrected the measurement results of ULCS which were measured by Keithley 6430. DC capacitance value was calculated from (1), by using the corrected current value and voltage ramp value.

$$C = \frac{I}{dV/_{dt}} \qquad (1)$$

In the measurements one period consists of four consecutive sections: zero(offset) current, positive current, zero(offset) current and negative current. Net current values for positive and negative current values were calculated by subtracting zero (offset) current value from positive and negative current readings respectively.

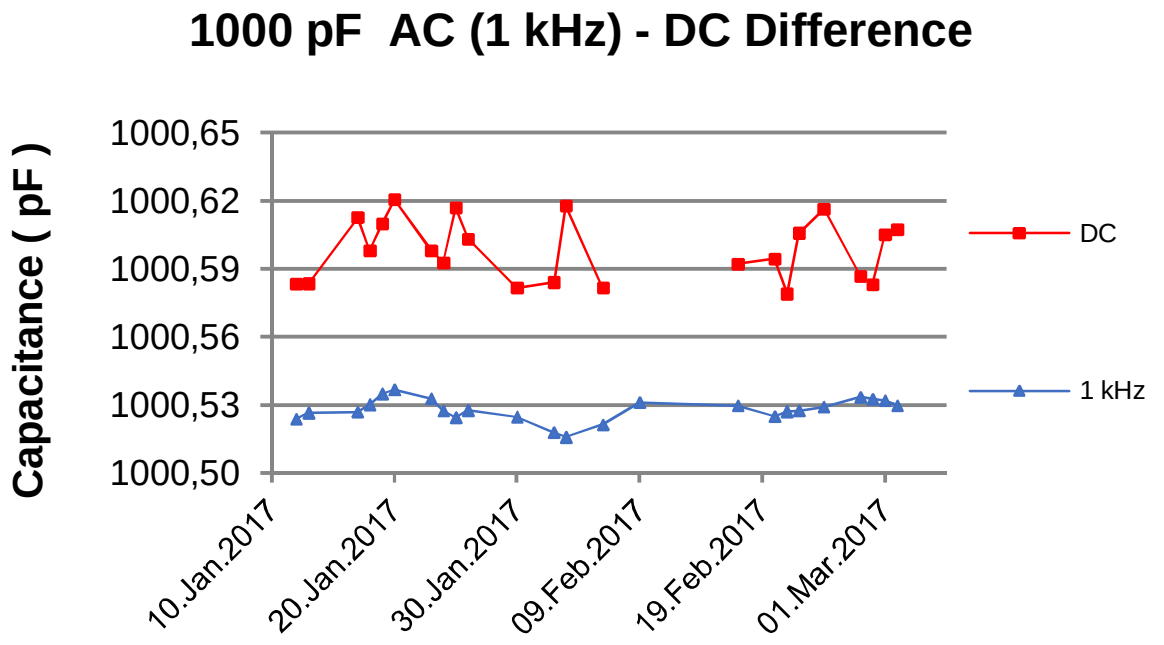

Fig. 8. 1000 pF AC-DC measurement results.

1000 pF measurements are shown in Fig. 8. AC-DC capacitance difference was found as 70 µF/F with a relative expanded uncertainty of 50 µF/F (*k=2*). The expanded uncertainty contributions (*k=2*) for the measurements of 1000 pF AC-DC capacitance difference are as follows:

- Uncertainty of producing 100 pA DC current by voltage-resistance method: 25 µA/A
- Ramp voltage uncertainty (100 mV/s): 30 µV/V
- Repeatability: 30 µF/F

We used 0.1 V and a 10 GΩ standard resistor to generate 10 pA DC current. 100 pF capacitor and 100 mV/s ramp voltage were used in the ULCS to obtain 10 pA current. Using the same method described above, the 100 pF AC-DC value difference was found as 105 µF/F with relative expanded uncertainty of 77 µF/F (*k*=*2*). The expanded uncertainty contributions (*k*=*2*) for the measurements of 100 pF AC-DC capacitance difference are as follows:

- Uncertainty of producing 10 pA DC current by voltage-resistance method: 55 µA/A
- Ramp voltage uncertainty (100 mV/s): 30 µV/V
- Repeatability: 44 µF/F

AC-DC capacitance difference measurement results and uncertainties are summarized in Table I.

TABLE I
SUMMARY TABLE FOR AC-DC CAPACITANCE MEASUREMENTS

| Capacitance Value | ULCS Settings | Voltage-Resistance Method Settings | AC-DC Capacitance Difference | Uncertainty ( k=2 ) |
|---|---|---|---|---|
| 1000 pF | 1000 pF<br>100 mV/s | 0.1 V<br>1 GΩ | 70 µF/F | 50 µF/F |
| 100 pF | 100 pF<br>100 mV/s | 0.1 V<br>10 GΩ | 105 µF/F | 77 µF/F |

### *F. Voltage Ramp Generator*

We used a NI-USB 4431 data acquisition (DAQ) card capable of supplying ±3.3 V to generate the voltage ramp. The DAQ card was programmed to produce any ramp voltages with slopes between 1 mV/s and 100 mV/s with 0.1 mV/s resolution. It is possible to obtain the desired voltage ramp slopes at different voltage levels. We obtained the same voltage ramp using different voltage levels and compared the voltage ramp stability. According to these measurement results we defined the optimum values that give the best stability. Sample rates of the DAQ card was another parameter for the generation of ramp voltage. DAQ card sample rates were optimized to give the best stability performance. The ramp voltage in one period was programmed in such a way that it was held at minimum voltage value first, ramped up to maximum voltage value then, held at maximum voltage value and ramped down to minimum voltage level at the end to eliminate the offset currents in the current measurement devices.

Characterization of the DAQ card was performed with a calibrated HP 3458A multimeter using an external time base of 1 s. Time base signal was produced from microprocessor card which has a stable 10 MHz crystal oscillator. Short time stability of the time base is less than 0.2 µs/s in an hour and annual drift is less than 2 µs/s. The multimeter is programmed to perform measurements using external trigger function.

Characterization measurements of voltage ramps were performed periodically in decadic steps. Measurements were performed as follows: From 1 mV/s to 10 mV/s with 1 mV/s steps, from 10 mV/s to 100 mV/s with 10 mV/s steps. The ramp voltages which are not decade value were estimated using linear fit equations for two different ranges of 1 to 10 mV/s and 10 to 100 mV/s. The linear fit equation uncertainties used for estimating the non-decadic voltage ramp values are less than 20 µV/V. The ramp voltage behaviors are given in Fig. 9 to Fig. 11. Standard deviation of the 10 mV/s voltage ramp is about 30 µV/V for nearly 2 years. The fluctuations seen in these figures usually occur when the device is switched on after the device was turned off for long holiday periods.

We also checked for short time stability and non-linearity of the voltage ramp generator for 1 mV/s, 10 mV/s and 100 mV/s voltage ramp values and found them to be less than 800, 80 and 8 µV/V respectively.

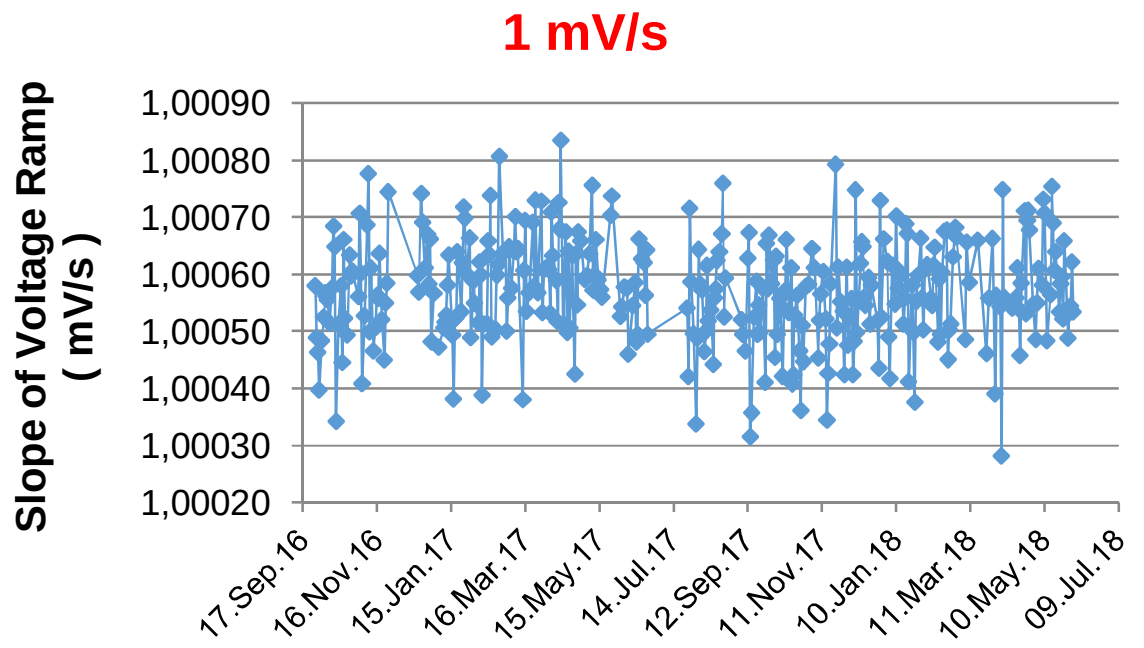

Fig. 9. 1 mV/s ramp voltage measurement results versus time.

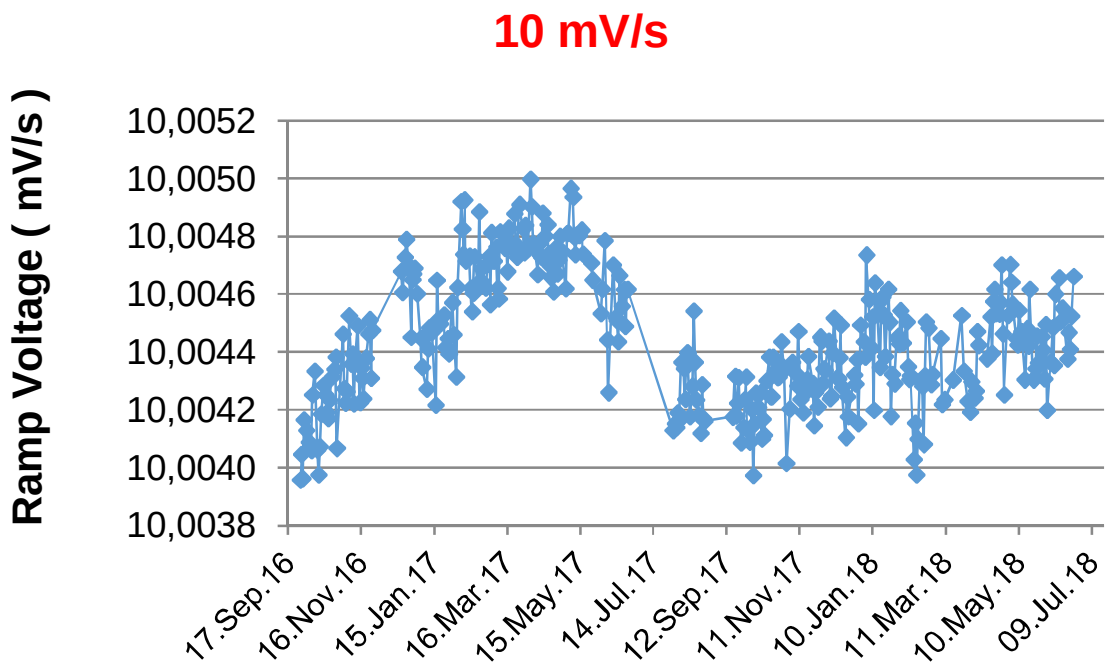

Fig. 10. 10 mV/s ramp voltage measurement results versus time.

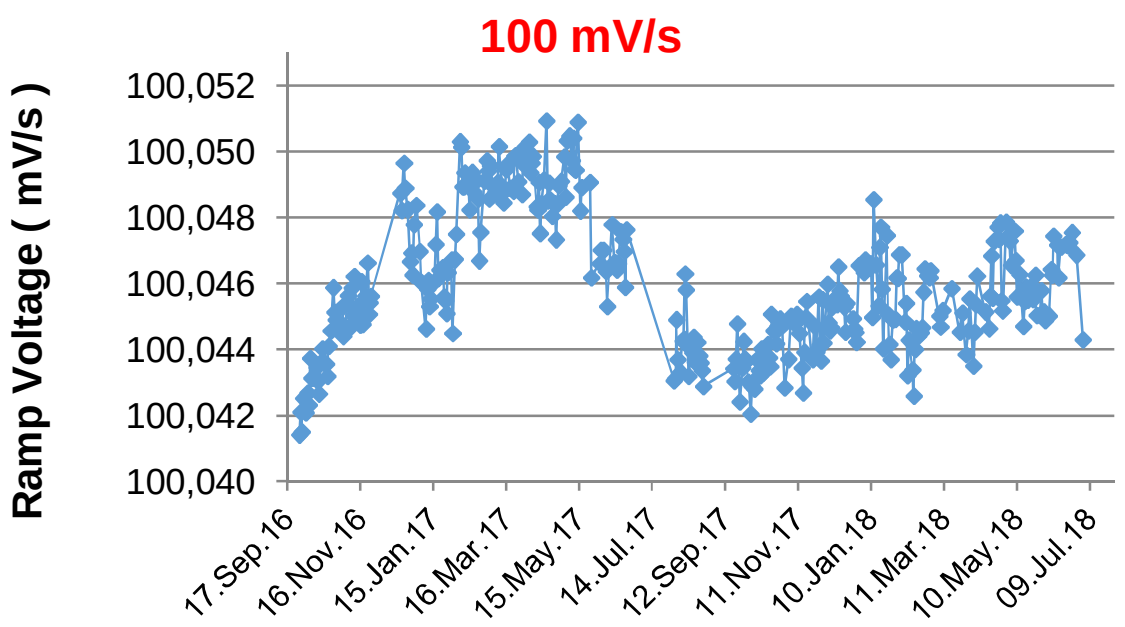

Fig. 11. 100 mV/s ramp voltage measurement results versus time.

It is possible to obtain the same current value using different capacitors and ramp voltages. Thus, it is important to determine optimum capacitance and ramp voltage values for each current value. We calibrated a Keithley 6430 using ULCS in different capacitor and voltage ramp combinations. The observed stabilities with Keithley 6430 is given in Table II.

TABLE II
CURRENT STABILITIES AT DIFFERENT CAPACITANCE AND VOLTAGE RAMPS

| Current | Capacitance ( pF ) | Voltage Ramp ( mV/s ) | Stability ( µA/A ) |
|---|---|---|---|
| 10 pA | 1000 | 10 | 110 |
| | **100** | **100** | **38** |
| 1 pA | 1000 | 1 | 2500 |
| | **100** | **10** | **120** |
| | 10 | 100 | 400 |
| 100 fA | 100 | 1 | 2000 |
| | **10** | **10** | **600** |
| | 1 | 100 | 1125 |
| 10 fA | **10** | **1** | **6000** |
| | 1 | 10 | 8000 |

The stability parameter in Table II is the standard deviation of 10 measurement periods (each period consisting of four sections; zero-positive-zero-negative). In all sections, only the values from 40% to 90% part of a section were evaluated. The time for each section were chosen as follows: 100 s for 1 mV/s and 10 mV/s voltage ramps, 60 s for 100 mV/s voltage ramp. Optimum capacitance and voltage ramp values that give the best stability with Keithley 6430 were chosen. The selected values were highlighted in the Table II. There is only one possible capacitance and voltage ramp value to produce 100 pA and 1 fA. 1000 pF and 100 mV/s were selected for 100 pA, 1 pF and 1 mV/s were selected for 1 fA.

## III. ULCS DEVICE

The left (right) side of Fig. 12 shows the front (rear) panel of the ULCS device. ULCS was manufactured to fit directly 19″ rack cabinets. The ULCS is fully computer controlled via USB connection.

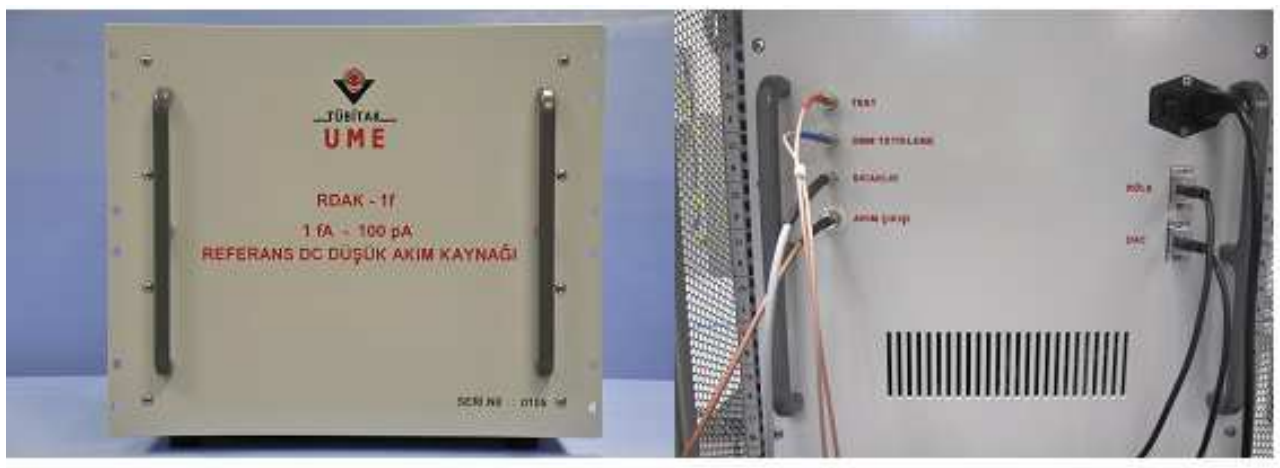

Fig. 12. The ULCS device front and back view.

Connections are at the rear panel of the device. There are current output, ramp voltage output, external time base output and temperature output. All these outputs are BNC type connectors except temperature output which is 4 pin LEMO connector.

The software user interface provides routines for the current source, ramp voltage measurement, Keithley 6430 and 6517 type electrometer calibration menus. In the current source menu, any desired current can be adjusted with 0.1 fA resolution. Offset (zero) current can also be applied from the ULCS device.

## IV. UNCERTAINTY

Main uncertainty components for ULCS current source are as follows: Capacitor, ramp voltage and time base. Table III shows the uncertainties for the current generated by the ULCS. The values are type B uncertainties whereas type A uncertainty is dependent on the specific measurement device.

The type B uncertainty parameters are explained as follows:

Capacitor:

- Calibration with AH2500A Capacitance Bridge (1 to 1000 pF) : 10 µF/F at 1 kHz.
- Temperature dependence of capacitors (1 to 1000 pF): 0.5 µF/F for 10 mK temperature change.
- AC-DC frequency dependence for 1 and 10 pF: 150 µF/F estimated from measurements between 50 Hz to 1 kHz with AH 2700A Capacitance Bridge.
- DC capacitance measurement uncertainty for 100 pF and 1000 pF 77 µF/F and 50 µF/F respectively. DC capacitance measurement details are explained in Section E.

Time base:

- Calibration with frequency counter traceable to TUBITAK UME national time standards: 2 µs/s (including drift of time base)

Ramp Voltage Generator:

- Calibration with HP 3458A multimeter: 10 µV/V
- Uncertainty of ramp voltage due to stability and nonlinearity for one measurement set:
  - For 1 mV/s ramp voltage: 1200 µV/V
  - For 10 mV/s ramp voltage: 120 µV/V
  - For 100 mV/s ramp voltage: 12 µV/V

TABLE III
UNCERTAINTY OF ULCS

THE VALUES ARE ONLY TYPE B UNCERTAINTIES WHEREAS TYPE A UNCERTAINTIES ARE DEPENDENT ON THE SPECIFIC MEASUREMENT DEVICE

| $I$ | $U(I)$ ( $k$=2 ) |
|---|---|
| **1 fA** | 2.5 aA ( $2.5 \cdot 10^{-3}$) |
| **10 fA** | 25 aA ( $2.5 \cdot 10^{-3}$) |
| **100 fA** | 40 aA ( $4 \cdot 10^{-4}$) |
| **1 pA** | 0.3 fA ( $3 \cdot 10^{-4}$) |
| **10 pA** | 1 fA ( $1 \cdot 10^{-4}$) |
| **100 pA** | 6 fA ( $6 \cdot 10^{-5}$) |

## V. MEASUREMENT RESULTS

We performed calibration of Keithley 6430 electrometer at different current levels. 95 pA nominal current results are shown in Fig. 13. Each measurement period consists of four sections: 60 s of zero (offset) current, 60 s of positive current, 60 s of zero (offset) current and 60 s of negative current. For each section of a period, the current values from 40% to 90% part of that section were evaluated. The rest of the currents in the same section were discarded because of current stabilization and polarity reversal settlings. The discarded data are not shown in the graph for simplicity. Fig. 13 shows data for 10 measurement periods.

"+95 pA", "-95 pA" and "Offset 95 pA" sections in Fig. 13 correspond to +95 pA, -95 pA and zero current values applied from ULCS respectively. Measurement mean values and standard deviations of those sections obtained by Keithley 6430 are -95.127 pA with 70 µA/A standard deviation, 95.126 pA with 65 µA/A standard deviation and 0.002 pA with 0.01 pA standard deviation respectively.

The uncertainty of the ULCS for 100 pA range is 60 µA/A. There are fluctuations approximately 150 µA/A in the measurement results which can be seen in Fig.13 with red bars in "+95 pA" section. These fluctuations arise from Keithley 6430 itself.

We also performed measurements with standard version of Ultra stable Low-noise Current Amplifier (ULCA) from Physikalisch-Technische Bundesanstalt (PTB) Germany [9], [10] in the scope of the Joint Research Project 'e-SI-Amp' (15SIB08). Evaluation of the measurements is still in progress.

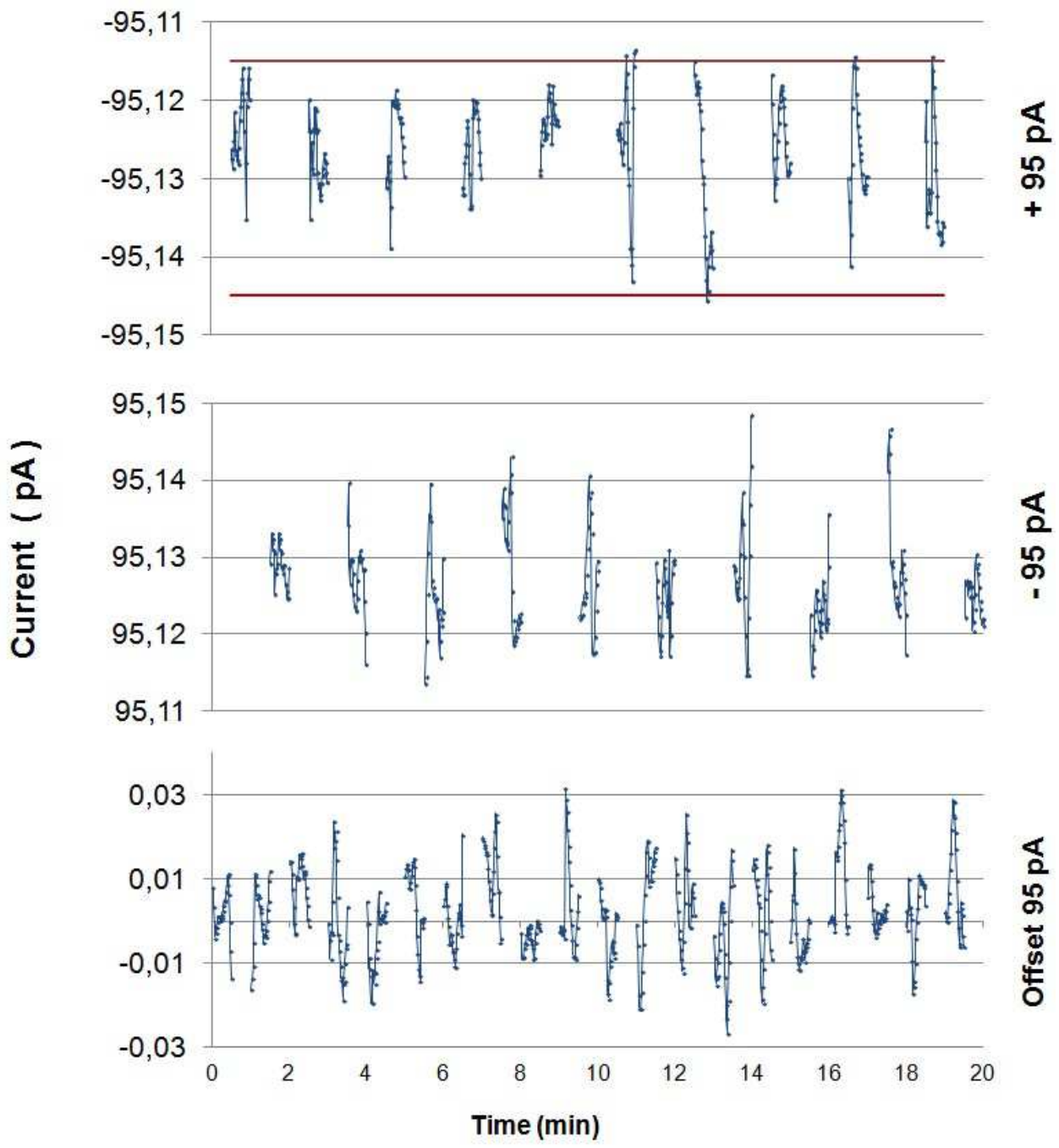

Fig. 13. ULCS measurement results of Keithley 6430 calibration at 95 pA current.

## VI. Conclusion

We developed an ultra-low DC current source (ULCS) in the range of 1 fA-100 pA with ultra-low resolution of 100 aA. Preliminary results show that uncertainties are approximately 2.5 mA/A and 0.1 mA/A at currents of 1 fA and 100 pA respectively.

We preferred to use a commercial D/A card which has 24 bit resolution from NI (National Instruments) to generate voltage ramp. In our method we did not make any corrections to the output of the D/A convertor.

The device is suitable for the low current applications such as optical detector characterizations, night vision systems and medical instrumentation systems. ULCS is capable of automatic calibration of Keithley 6430 and 6517 model electrometers up to 100 pA range.

Further investigations will be performed on stability of ramp voltage generation and on loss factor improvement of 1 pF capacitance value.

## Acknowledgment

The authors would like to thank Turgay Özkan for his assistance in relay card and microprocessor software development.

This work was supported by the Joint Research Project ‘eSI-Amp’ under Grant 15SIB08. This project has received funding from the European Metrology Programme for Innovation and Research (EMPIR) cofinanced by the Participating States and from the European Union’s Horizon 2020 Research and Innovation Programme.

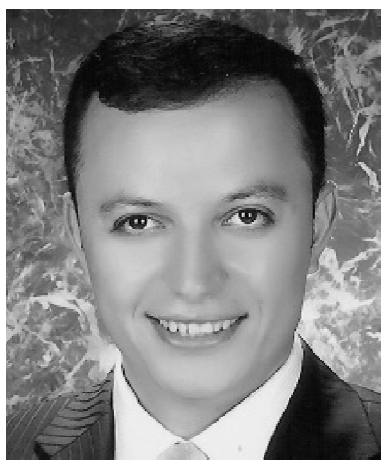

**Ömer ERKAN** was born in Bozkır, Konya, Turkey in 1979. He received the B.S. degrees in Electrics & Electronics Engineering in 1999 and M.Sc. degrees in Electronics Engineering in 2003 from the Istanbul University.

From 1999 to 2002, he worked as research assistant at the Istanbul University Electric & Electronics Engineering department where he involved coding techniques in wireless communication systems. Since 2002, he has been working as Chief Senior Researcher at Impedance Laboratory in National Metrology Institute of Turkey (TUBİTAK UME).

He has graduated with the first honor degree from Istanbul University Electric & Electronics Engineering department in 1999.

His research interests include DC resistance metrology, impedance metrology, development of primary level metrological capacitance, inductance, AC/DC resistance standards and bridges, development of DC resistance, capacitance, inductance standards, and development of ultra-low DC current measurements systems.

Mr. Erkan is Assessor in the field of Electrical Metrology for Turkish Accreditation Agency (TURKAK) since 2008.

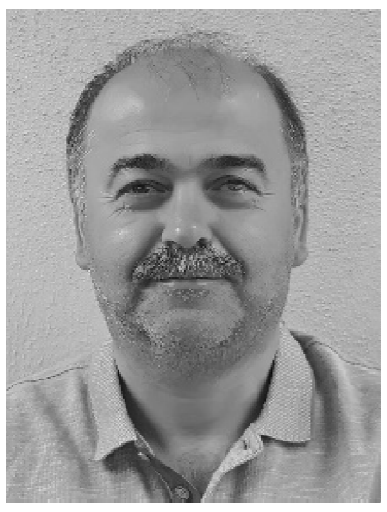

**Yakup GÜLMEZ** was born in, Bursa, Turkey in 1969. He received the B.S. degree in Electronics and Communication Engineering from Yıldız Technical University İstanbul in 1990. He received the M. S. degree in Business and Administration from Yıldız Technical University İstanbul in 1993.

He was a Senior Researcher at Impedance Laboratory of National Metrology Institute of Turkey (TUBİTAK UME) between 1993 and 2016. Since September 2016, he has been Head of Analytic Device and Systems Department of National Electronic and Cryptology Institute of Turkey (TÜBİTAK BİLGEM). He is author of more than 20 articles and more than 30 technical reports. His research interests include impedance metrology, development of primary level capacitance, inductance, AC resistance measurement bridges, development of capacitance, inductance, AC resistance standards, inductive voltage dividers, DC resistance standards, development of methods for impedance measurements. His research interest also includes designing devices for various electrochemical applications, development of wide range of optical spectroscopic device and systems.

Mr. Gülmez is technical electrical metrology expert for Turkish Accreditation Agency (TURKAK).

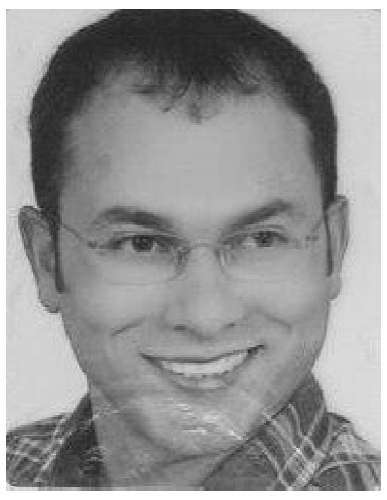

**Cem HAYIRLI** was born in Ankara, Turkey in 1974. He received the B.S. degrees in physics in Middle East Technical University in 1998 and M.S. degrees in physics from the Kocaeli University in 2007

From 1998 to 2012, he worked as researcher at RF & Microwave Laboratory in National Metrology Institute of Turkey (TUBİTAK UME) and has been working at Impedance Laboratory at the same institute since 2012. His research interests include resistance, inductance and capacitance measurement methods.

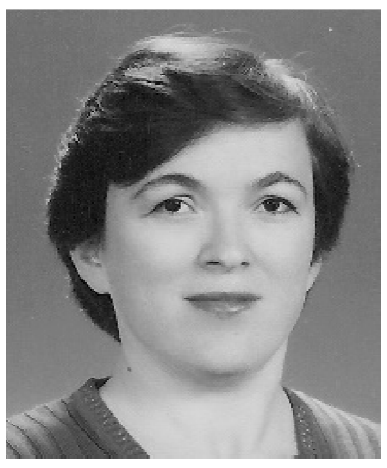

**Gülay GÜLMEZ** was born in Karaman, Turkey in 1969. She received the B.S. degree in Electric and Electronics Engineering from the University of Hacettepe, Ankara, in 1992.

Since 1992, she has been a Senior Researcher at Impedance Laboratory of National Metrology Institute of Turkey (TUBİTAK UME). She is author of more than 20 articles and more than 30 technical reports. Her research interests include impedance metrology, development of primary level capacitance, inductance, AC resistance measurement bridges, development of capacitance, inductance, AC resistance standards, inductive voltage dividers, DC resistance standards, development of methods for impedance measurements. Mrs. Gülmez is EURAMET TCEM technical reviewer for impedance metrology and Assessor for Turkish Accreditation Agency (TURKAK) since 2005.

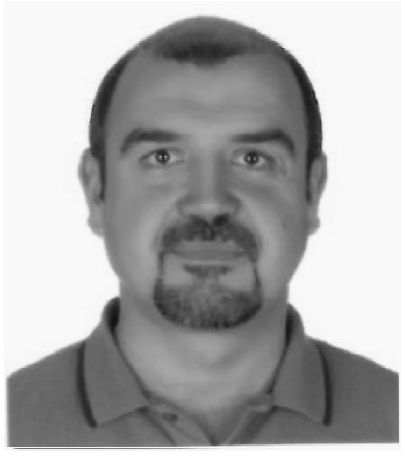

**Enis TURHAN** was born in Balıkesir, Turkey in 1975. He received the B.S. degree in Electronics and Communication Engineering in 1996 from Istanbul Technical University.

Since 1997, he has been working as senior researcher at Impedance Laboratory of National Metrology Institute of Turkey (TUBİTAK UME).

His research interests include inductance and capacitance measurements, inductive voltage dividers, coaxial AC bridges, primary DC resistance, ultra-low DC current measurements.

Mr. Turhan is Assessor in the field of Electrical Metrology for Turkish Accreditation Agency (TURKAK) since 2008.